\documentstyle[aps,pre,multicol,epsf]{revtex}

\begin{document}

\draft

\newcommand {\qea}  {\mbox{$q_{\scriptscriptstyle {\rm EA}}$}}

\title{
Ultrametricity in 3$d$ Edwards-Anderson spin glasses
}
\author{
Silvio Franz and
Federico Ricci-Tersenghi
}
\address{
Abdus Salam International Center for Theoretical Physics\\
Strada Costiera 11, P.O. Box 586, 34100 Trieste (Italy)
}
\date{\today}
\maketitle

\begin{abstract}
We perform an accurate test of Ultrametricity in the aging dynamics of
the three dimensional Edwards-Anderson spin glass. Our method consists
in considering the evolution in parallel of two identical systems
constrained to have fixed overlap. This turns out to be a particularly
efficient way to study the geometrical relations between
configurations at distant large times. Our findings strongly hint
towards dynamical ultrametricity in spin glasses, while this is absent
in simpler aging systems with domain growth dynamics. A recently
developed theory of linear response in glassy systems allows to infer
that dynamical ultrametricity implies the same property at the level
of equilibrium states.
\end{abstract}

\begin{multicols}{2}
\narrowtext

The dispute about the nature of the spin glass phase of finite
dimensional spin glasses has lasted by now almost twenty years, after
that the droplet model~\cite{DROPLET} has challenged the predictions
of mean field theory (MFT)~\cite{MEPAVI}.

In the last years a large collection of numerical data have tested,
with positive answer, the applicability of spin glass mean field
picture to finite dimensional ($d$=3,4)
systems~\cite{ROMANI,FDT,MAPARU}.  However some of these numerical
evidences have been recently re-interpreted as finite volume effects
by the authors of Ref.~\cite{DROSSEL} and so many questions on the
low-temperature phase of finite dimensional spin glasses still remain
unaswered.

One of the most characteristic aspects of mean field theory is the
prediction of ultrametricity (UM)~\cite{MPSTV}. At low temperature the
ergodicity is broken and many low free energy states are
present. Ultrametricity implies that the distances between these
states verify an inequality stronger than the triangular one (see
below). However this property has been rather elusive to a direct
probe.  The best evidences in favor of this property have been given
(to our knowledge) in Refs.~\cite{ULTRA,HART}.  In the former it was
studied the equilibrium of a 4$d$ spin glass model using small
samples.  In the latter very-low-energy configurations of a 3$d$ model
were used, again for relatively small samples.  In both cases an
extrapolation to large volumes was needed in order to check
ultrametricity.  In this paper we present results for the 3$d$ spin
glass model with a method that allows us to reach much larger sizes.

Another fundamental aspect of MFT is the prediction of slow dynamics
and aging~\cite{CUKU,CUKUSK,FRAME}. This is a non stationary
asymptotic regime, following a quench from a high temperature, which
persists forever in infinite systems.  In this out-of-equilibrium
regime, the equilibrium property of ultrametricity has a dynamical
counterpart in ultrametric relations among time dependent
autocorrelation functions.  Recent results of linear response theory
succeeded to relate in a unique way properties of statics and
dynamics, relying on the recently introduced hypothesis of stochastic
stability~\cite{FRAMEPAPE}. This property states the continuity of the
average correlation functions under weak random perturbation of the
Hamiltonian. The validity of the property in three and four
dimensional spin glasses has been numerically verified in
Ref.~\cite{FDT}. A remarkable consequence of stochastic stability is
that dynamical ultrametricity implies the static
one~\cite{FRAMEPAPE2}.

In this paper we investigate the possibility for dynamical UM in the
three dimensional Edwards-Anderson (EA) model. For comparison we also
perform the same test on models with domain coarsening off-equilibrium
dynamics, where UM should not be expected.

The high evidence we can achieve is based on a new dynamical method
where we evolve in parallel two identical systems (replicas for short)
with fixed value of the mutual overlap. This is analogous to a
conserved-order-parameter dynamics in a ferromagnetic system and it is
similar to the one already used in Ref.~\cite{ULTRA} to study the
equilibrium behavior.

The EA model is defined by the Hamiltonian $H({\bf S})=\sum_{<i,j>}
J_{ij} S_i S_j$ where the spins are Ising variables, the sum spans the
nearest neighbors pairs on a cubic lattice, and the couplings $J_{ij}$
are normally distributed quenched independent random variables.  We
define the overlap among two spin configurations ${\bf S}^1$ and ${\bf
S}^2$ as $q_{12}=L^{-3}\sum_{i} S^1_i S^2_i$, which is directly
related to the Hamming distance $d_{12}$ through $q_{12}=1-d_{12}$.

The mean field equilibrium solution of this model has the ultrametric
property at low temperature; for each three configuration ${\bf S}^1$,
${\bf S}^2$ and ${\bf S}^3$ chosen with Boltzmann probability, the
following inequalities hold:
\begin{equation}
q_{12} \geq \min\{ q_{13},q_{23} \} \quad , \quad
d_{12} \leq \max\{ d_{13},d_{23} \} \quad ,
\label{sum}
\end{equation}
which are much stronger than the usual triangular one $d_{12}\leq
d_{13}+ d_{23}$.  Moreover in Ref.~\cite{INPARU} has been shown that,
if some kind of ultrametricity holds in the low-temperature phase of
the EA model, it must be of the same kind of that present in the
mean-field solution.

In the off-equilibrium dynamical solution a relation analogous to
Eq.(\ref{sum}) holds for the two-time auto-correlation functions:
$C(t,t')=L^{-3}\sum_{i} S_i(t) S_i(t')$.  Taken three large times
$t_1\ll t_2\ll t_3$, one finds that
\begin{equation}
C(t_1,t_3)= \min \{ C(t_1,t_2),C(t_2,t_3) \}
\label{dum}
\end{equation}
The precise statement is that the relation among the correlations {\it
should tend} to the one of Eq.(\ref{dum}) in the infinite time
limit~\cite{CUKUSK,FRAME}. In the simulations the relation among the
three correlations is plagued by strong finite time effect, so that a
direct verification would be difficult with the present computer
resources.  We therefore decided to probe a relation which for long
times is consequence of, and equivalent to, Eq.(\ref{dum}).  We
consider the dynamics of two replicas of the system with the same
disorder, ${\bf S}^1$ and ${\bf S}^2$, which evolve in parallel with
different thermal noises and constrained at each time to have a fixed
mutual overlap $q_0$.  The dynamics follows a quench at time zero and
we measured the auto- and cross-correlation functions
\begin{eqnarray}
C(t,t')&=&\frac{1}{L^3}\sum_{i} S_i^1(t)
S_i^1(t')=\frac{1}{L^3}\sum_{i} S_i^2(t) S^2_i(t') , \nonumber\\
D(t,t')&=&\frac{1}{L^3}\sum_{i} S_i^1(t)
S_i^2(t')=\frac{1}{L^3}\sum_{i} S_i^2(t) S^1_i(t') .
\end{eqnarray}
Note that $C(t,t)=1$ for Ising variables, while the constraint implies
$D(t,t)=q_0$.

We would like to argue, with a hand waving argument, that the relation
in Eq.(\ref{dum}) entails the following ultrametric constraint on the
cross-correlation function if the value of $q_0$ is between $0$ and
the value of the Edwards-Anderson parameter \qea:
\begin{equation}
D(t,t')= \min \{ C(t,t'),q_0 \}
\label{cum}
\end{equation}
During the relaxation the free energy of the system decreases
monotonically towards its equilibrium value. This will be higher or
equal to the one of the unconstrained system at the same temperature.
The basic observation is that, if $q_0$ is one of the values allowed
for the overlap among equilibrium states, {\it i.e.}\ $q_0\in
[0,\qea]$, the equilibrium free-energy of the constrained system
should coincide with the one of the unconstrained
one~\cite{FPV}. Eq.(\ref{dum}) expresses the fact that for large
times, the directions in which the system can go without increasing
the free-energy must be compatible with ultrametricity. The
constrained system can lower its free-energy down to the equilibrium
value only if all the possible correlations one can form (auto and
cross) verify for long times UM inequalities. As two of the 4
correlations that can be formed with the configurations ${\bf
S}^1(t)$, ${\bf S}^1(t')$, ${\bf S}^2(t)$ and ${\bf S}^2(t')$ are
fixed to $q_0$, Eq.(\ref{cum}) should follow.  A formal argument
leading to the same conclusions can be formulated~\cite{NOI-PREP},
extending to constrained systems the correspondence among statics and
dynamics mentioned above.

\begin{figure}
\epsfxsize=0.9\columnwidth
\epsffile{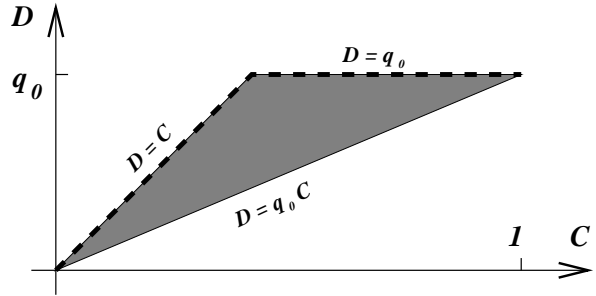}
\caption{Schematic draw of the region of allowed values for $C$ and
$D$ (shaded area). The ultrametric relation force them on the bold
dashed line.}
\label{area}
\end{figure}

In order to see how restrictive relation in Eq.(\ref{cum}) is, let us
discuss what one can expect on a general ground for the relation among
$D$ and $C$ in a relaxational system. In Fig.~\ref{area} we display
the set of allowed values for $C(t,t')$ and $D(t,t')$ (shaded area),
simply assuming the triangular relation and a monotonic decrease of
both functions when the time argument's difference increases.  The set
of values allowed by the UM relation is represented with the bold
dashed line.

\begin{figure}
\epsfxsize=0.9\columnwidth
\epsffile{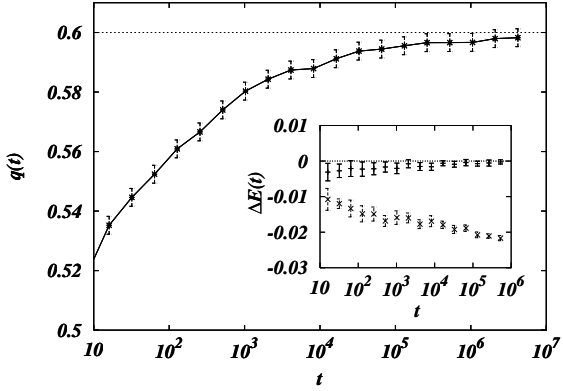}
\caption{The time evolution of the overlap between the pairs of EA
systems at temperature $T=0.7$ with $q_0=0.6$.  In the inset we show
the energy difference between a constrained and a free system at the
same temperature.  The upper data correspond to $T=0.7$ and
$q_0=0.6<\qea$ and are clearly compatible with zero.  The lower data
correspond to $T=0.9$ and $q_0=0.9>\qea$ and show how the energy of
the constrained system is lower.}
\label{conv}
\end{figure}

In order to check whether Eq.(\ref{cum}) holds in the 3$d$ EA model,
we have simulated, by Monte Carlo method (Metropolis update), two
coupled systems with a soft constraint $\sum_i S_i S_i' \approx L^3
q_0$, imposed modifying the Boltzmann weight with a Gaussian of width
proportional to $L^{-3/2}$, such that the weight of a configuration
would be $\exp[-\beta(H({\bf S}_1)+H({\bf S}_2))-\lambda L^3
(q_{12}-q_0)^2/2]$.  The value of the $\lambda$ parameter must be
appropriately tuned: a too small value would not force enough the
systems and their overlap will be systematically different from the
one we fixed ($q_0$).  On the other hand a too large value would
render movements too unprobable and the dynamics would evolve very
slowly. In the whole set of runs we fixed $\lambda=5$ for the EA model
and $\lambda=2$ for the other models.  The choice has been made with
the aim of maximizing the Monte Carlo acceptance rate, avoiding the
systematic errors just described.  We have checked (see
Fig.~\ref{conv}) that, for these values of $\lambda$, $q_{12}(t)$
tends to the desired value $q_0$.

As we study the behavior of the model in the aging regime, we do not
need to reach thermalization and we can simulate large sizes
($L\!=\!24$). For such volume, finite size effects do not affect the
dynamical regime we study.  The critical temperature of a single
uncoupled model is $T_c=0.95(4)$~\cite{MAPARU} and we simulate the
system in the spin glass phase at $T=0.7$.  The large size, the low
temperature and the starting configuration, which is randomly chosen,
ensure that the system stays in the aging regime all along the
simulation which can be as long as $10^8$ MCS.  These high
performances have been achieved using the parallel computer
APE100~\cite{APE}.

All the correlation functions we measure are extensive and
self-averaging quantities, so we do not need a large number of
disorder realizations.  Their fluctuations are small thanks to the
large volume used and we average the results on a quite small number
of samples ($N_S=10$). The error on the data is always calculated as
the sample-to-sample fluctuation.

The first quantities we have studied are the overlap of the two
replicas and the internal energy as a function of time.  As announced,
we see in Fig.~\ref{conv} that the overlap converges towards the value
$q_0$ we have fixed.  The ``soft'' way of imposing the constraint is
evident in the behavior of the overlap, which is not equal to $q_0$
during all the run, but clearly converges to it.  In the forecoming
analysis we will use only the data obtained in the time range where
the overlap is statistically compatible with $q_0$, that is $t_w \geq
10^5$.  The data in the inset of Fig.~\ref{conv} show the difference
between the energy of the constrained system and that of a free system
evolving at the same temperature after a quench at time zero. As we
expected, the energy difference tends to zero, as it should if the
value $q_0$ is allowed at equilibrium, while for $q_0>\qea$ the energy
difference goes to a finite value.

\begin{figure}
\epsfxsize=0.9\columnwidth
\epsffile{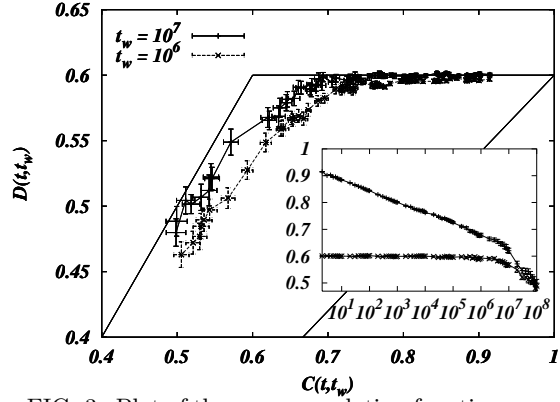}
\caption{Plot of the cross-correlation function versus the
autocorrelation one in the 3$d$ EA model at temperature $T=0.7$ for
two waiting times $t_w=10^6,10^7$. The curves approach the UM bound as
$t_w$ grows. The solid lines are the boundaries of the shaded area of
Fig.~\ref{area}.  In the inset we display the time dependence of the
two functions.}
\label{sg}
\end{figure}

In Fig.~\ref{sg} we show our main result: plotting the
cross-correlation as a function of the auto-correlation we obtain, for
relatively long times, that the data points are quite close to the UM
bound.  A perfectly ultrametric system would stay on the line
$D=q_0=0.6$ as long as $C \geq q_0$ and then would follow the line
$D=C$ (both lines are plotted in Fig.~\ref{sg}).  The data for the 3$d$
EA model are clearly converging to the UM bound: we remind the reader
that the data cannot leave the shaded area of Fig.~\ref{area}, that is
they cannot cross any of the boundary lines reported in Fig.~\ref{sg},
and so we expect that they naturally converge to the UM bound.

To understand better how probing are our data, we have also simulated
two models where UM is not expected to hold: the 2$d$ ferromagnetic
Ising model, and its site diluted version in 3$d$. The off-equilibrium
dynamics of both models shows a domain coarsening regime where the
correlation function depends on both times as $C(t,t')\approx {\cal
C}(t'/t)$, a scaling form incompatible with ultrametricity.

We have simulated a 2$d$ pure ferromagnetic Ising model of linear size
$L=2000$ at a temperature $T=1.5$ ($T_c^{2d} \simeq 2.27$).  We choose
a very large size and small waiting times ($t_w=32,64,128$) in order
to avoid that the system gets out from the aging regime we are
interested in. These times are sufficiently large to be close to the
scaling regime. The data are averaged on $N_S=100$ different noise
realizations.

\begin{figure}
\epsfxsize=1.05\columnwidth
\epsffile{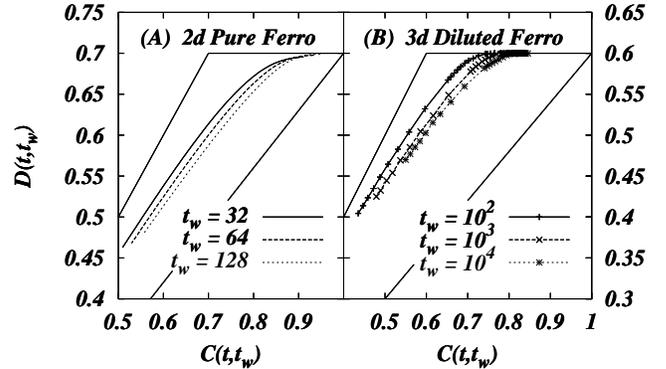}
\caption{The same plot as Fig.~\ref{sg} for (A) the pure ferromagnet
in 2$d$ at $T=1.5$ and (B) the 3$d$ diluted ferromagnet at
$T=1.67$. The waiting time effect is opposite to the one observed in
the spin glass.}
\label{ferro_dilu}
\end{figure}

The data from the pure ferromagnetic system are plotted in
Fig.~\ref{ferro_dilu}A in the usual $D(t,t_w)$ versus $C(t,t_w)$ plot.
We also report the bounds of the allowed region (the shaded region of
Fig.~\ref{area}).  Note that the data are far away from the UM bound
and they seem to converge to some $t_w$-independent curve.

Maybe one can think that the comparison of a spin glass with the pure
ferromagnet is not enough. So we have simulated also a 3$d$ site-diluted
ferromagnetic Ising model, which has a coarsening dynamics similar to
that of the pure model, but much slower~\cite{DYN_DILU} and
complicated by interface pinning.  We have simulated two samples (each
one consisting of a pair of interacting systems) of linear size
$L=200$ and spin concentration $c=0.65$.  The temperature is well deep
in the frozen phase, $T=1.67$, the critical temperature being
$T_c^{3d}(c=0.65) \simeq 2.70$~\cite{DILU_3D}.

In Fig.~\ref{ferro_dilu}B we see that the behavior of the data from
the diluted ferromagnetic model may resemble that of the EA model,
because it seems to be somehow close to the UM bound.

\begin{figure}
\epsfxsize=0.95\columnwidth
\epsffile{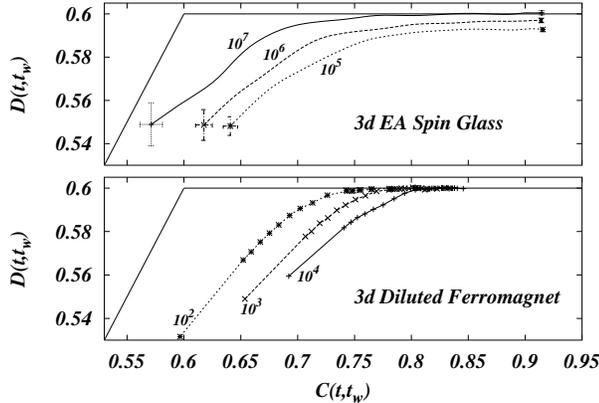}
\caption{Zooms of Figs.~\ref{sg} and~\ref{ferro_dilu}B in the region
of $D$ close to the upper boundary $D=q_0$. The comparison of the two
plots makes clear that the effect of the waiting time is opposite in
the two cases, and the long time extrapolation is certainly very
different.}
\label{zoom}
\end{figure}

Looking carefully at the figure we note, however, that the
$t_w$-dependence of the data in Fig.~\ref{sg} and in
Fig.~\ref{ferro_dilu}B are opposite.  In fact increasing the waiting
time, the value of the auto-correlation function at the point where
the data leave the horizontal line decreases in the spin glass case,
while it increases in the diluted ferromagnetic case.  To make this
effect clearer we zoomed the region of Fig.~\ref{sg}
and~\ref{ferro_dilu}B near the horizontal line (see Fig.~\ref{zoom}).

Summarizing, we have used a new numerical method to test
ultrametricity in short range spin glasses. We find evidence that for
long times the ultrametric equality between two time correlations
become fulfilled. As we already stressed the property of stochastic
stability implies then static ultrametricity. The behavior of spin
glasses is strikingly different from the behavior of ordered and
disordered models with domain coarsening, where we find
incompatibility with ultrametricity.

\section*{Acknowledgments}

We would like to thank Giorgio Parisi for continuous interactions and
advises, the ``INFN sezione di Roma I'' for the use of the APE100
computer and the ``Universit\`a di Roma La Sapienza'' for kind
hospitality.  We thank also M.~A.~Virasoro and R.~Zecchina for useful
discussions.

\end{multicols}
\end{document}